\documentclass[fleqn,usenatbib,useAMS]{mnras}

\usepackage{graphicx}	
\usepackage{amsmath}	
\usepackage{multicol}        
\usepackage{subcaption}
\usepackage{bm}		
\usepackage{pdflscape}	
\usepackage{wasysym}
\usepackage{enumitem}

\usepackage[T1]{fontenc}
\usepackage{ae,aecompl}

\usepackage{newtxtext,newtxmath}
\newcommand{\orcid}[1]{\href{https://orcid.org/#1}{\includegraphics[width=8pt]{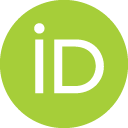}}}

\title[Fixed points]{Irregular Fixation: I. Fixed points and librating orbits of the Brown Hamiltonian}

\author[E. Grishin]{Evgeni Grishin$^{1,2}$ \orcid{0000-0001-7113-723X}%
\thanks{Contact e-mail: \href{evgeni.grishin@monash.edu}{evgeni.grishin@monash.edu}}%
\\
$^{1}$School of Physics and Astronomy, Monash University, Clayton, VIC 3800, Australia\\
$^{2}$OzGrav: Australian Research Council Centre of Excellence for Gravitational Wave Discovery, Clayton, VIC 3800, Australia}

\pubyear{2024}

\begin{document}

\label{firstpage}
\pagerange{\pageref{firstpage}--\pageref{lastpage}}
\maketitle

\begin{abstract}
In hierarchical triple systems, the inner binary is slowly perturbed by a distant companion, giving rise to large-scale oscillations in eccentricity and inclination, known as von-Zeipel-Lidov-Kozai (ZLK) oscillations. Stable systems with a mild hierarchy, where the period ratio is not too small, require an additional corrective term, known as the Brown Hamiltonian, to adequately account for their long-term evolution. Although the Brown Hamiltonian has been used to accurately describe the highly eccentric systems on circulating orbits where the periapse completes a complete revolution, the analysis near its elliptical fixed points had been overlooked. We derive analytically the modified fixed points including the Brown Hamiltonian and analyse its librating orbits (where the periapse motion is limited in range). We compare our result to the direct three-body integrations of millions of orbits and discuss the regimes of validity. We numerically discover the regions of orbital instability, allowed and forbidden librating zones with a complex, fractal, structure. The retrograde orbits, where the mutual inclination is $\iota > 90\ \rm deg$, are more stable and allowed to librate for larger areas of the parameter space. We find numerical fits for the librating-circulating boundary. Finally, we discuss the astrophysical implications for systems of satellites, stars and compact objects. In a companion paper (paper II), we apply our formalism to the orbits of irregular satellites around giant planets.
\end{abstract}





\section{Introduction} \label{sec:intro}

Triple and multiple systems are ubiquitous and appear in many astrophysical settings, from Solar-System asteroids \citep{marchis05}, to multiple planetary and stellar systems, \citep{martin14, wagner16}, to multiple stellar and compact objects \citep[where most massive stars are in triples,][]{offner23}, and supermassive black holes on galactic scales \citep{deane14}. The dynamical evolution of three bodies under the gravitational force -- the three body problem -- is a notoriously chaotic problem with no analytical solution \citep{poincre1892}. For strongly chaotic systems a statistical approach is generally preferred to yield a probabilistic outcome of the scattering in terms of the final states \citep{moh1, moh2}. Only recently, explicit dependence on the orbital elements had been found with various methods \citep{stone2019, ginat1, ginat2,kol1, kol2, Man2021}.

For hierarchical systems, the inner binary of semi-major axis $a_1$ is perturbed by a distant tertiary at semi-major axis $a_2 \gg a_1$, which is typically on a much longer orbit. This separation of timescales allows hierarchical systems to be stable. In addition, it is natural to expand the interaction Hamiltonian in power series in multipoles, where the semi-major axis ratio $a_1/a_2\ll 1$ is the small parameter. Truncating at the leading quadrupole ($\propto (a_1/a_2)^2$) order and orbit-averaging over both the inner and outer orbits \citep[e.g.][]{1harr1968} leads to a simplified problem where the typical timescale is a much longer, \textit{secular} timescale. An integrable,  analytical, approximate solution in the test-particle limit \citep[where one of the inner masses is small compared to the outer mass,][]{kinoshita07, lub21}\footnote{Formally, the condition is that the angular momentum of the inner binary is much smaller than the angular momentum of the outer binary, since the feedback on the outer binary is proportional to the angular momentum ratio. \cite{sod75} calls this approximation 'immovable wide orbit limit'.} can be obtained. A key feature is that for systems where the mutual inclination $\iota$ between the inner and outer orbital planes is large enough, the inner orbital eccentricity grows significantly, which is known as the von-Zeipel-Lidov-Kozai effect (ZLK) \citep{vZ1910,lid62,koz62}. Adding the higher order octupole ($\propto (a_1/a_2)^3$) terms render the system to chaotic again, and it is usually studied numerically by solving the secular equations of motion \citep[e.g.][]{naoz13}.

The latter double-averaging approximation was first tested when the apsidal and nodal precession of the Moon was inconsistent with the secular theory, which was off by a factor of $\sim 2$. Taking into account the relative motion of the Earth after one Lunar period and expanding for almost coplanar and circular orbits, the corrected longitude of pericentre advance 
$\dot \varpi_{\leftmoon}$ is \citep{brouwer61, tremaine23}
\begin{equation}
     \frac{d\varpi_{\leftmoon}}{d\tau} = \frac{3}{4}\epsilon + \frac{225}{32}\epsilon^2 + \frac{4071}{128}\epsilon^3 + \frac{265493}{2048}\epsilon^4 + \mathcal{O}\left(\epsilon^5\right)  \label{eq1}
\end{equation}
where $\tau=t/2\pi P_2$ is a normalised timescale, $P_1$ and $P_2$ are the inner and outer orbital periods, and $\epsilon = P_1/P_2 \approx 1/12$ is the period ratio. Thus the quadratic $\epsilon^2$ term is nearly equal to the linear $\epsilon$ term. 

While the Lunar theory had been developed for low inclination and eccentricity, discovering highly inclined and eccentric exoplanets and hundreds of irregular satellites required a more general development. Retrograde orbits have been observed to be more stable \citep{i79, i80, hb91}, while the exact dependence on the mutual inclination was studied relatively recently \citep{grishin2017, tory22}. The disturbing function for irregular satellites had been expanded in \cite{cb04}, and generalised by \cite{luo16} for any mass ratio where they find an 'effective potential'. \cite{tremaine23} had recently shown that the aforementioned (and other related) derivations are equivalent to the pioneering work of Brown \citep{brown1, brown2, brown3}, and coined the effective potential as 'Brown Hamiltonian'. The equivalence relies on an intricate gauge freedom in the averaging procedure. We note that it is also possible to evolve the 'single averaged' equations, where the outer period is the typical timescale \citep[e.g.][]{ll18}. {Generalised analogues
of Brown’s Hamiltonian for higher-order perturbations theory (but less practical) are given by \cite{beauge06, lei18, lei19}}.

The importance of the Brown Hamiltonian was highlighted in the context of highly eccentric orbits. \cite{gpf18} had developed analytical expressions for the modified maximal eccentricity and critical inclination, which were later extended beyond the test particle limit \citep{man22}.  Close approaches and disruptions in low-hierarchy triples were studied in \cite{haim18} and \cite{mushkin20}, respectively, while extreme eccentricities have been invoked in white dwarf collision rates \citep{kushnir13}. The modified formulae were widely used in many applications, from Kuiber-Belt collisions \citep{gri20, pc20}, to compact object mergers and gravitational waves \citep{fg19}. Other external driving forces such as the cluster and the galactic tidal potential for wide binaries and triples had been also recently studied \citep{ham19, bub20, gp22}.

For the {double-averaging} approximation ($\epsilon=0$), for initial eccentricity $e_0$, fixed points in the eccentricity $e_1$ -- argument of pericentre $\omega_1$ phase space appear (for allowed mutual inclination values, $\cos^2 \iota \le 3/5$) at $\omega_1=\pi/2$ and $e_1^2 = 1 - \sqrt{ (5/3)(1-e_0^2)}\cos \iota$ \citep{car02, antognini15, hamers-zlk}\footnote{Both \cite{car02} and \cite{antognini15} use the notation $\Theta = (1-e^2) \cos^2I$, while we'll write it it terms of $j_z = \sqrt{\Theta}$.}. Fixed points in the planetary limit were studied by \cite{hn20}. \cite{kk23} studied the dynamics of a ZLK cycle near a fixed point for a slowly precessing outer orbit and later generalised to libration far from the fixed point in \cite{kk24}.

Besides the highly eccentric cases, the analytical properties of the Brown Hamiltonian were overlooked. 
In this paper, we explore the fixed point and the phase space of the Brown Hamiltonian. We explore the regions where libration is allowed and uncover its complex, fractal structure and find numerical fit for the boundary far enough from instability. In a companion paper (paper II) we will focus on the motion of irregular satellites in the Solar system.

The paper is organised as follows: In sec. \ref{s2} we derive the equations of motion of the Brown Hamiltonian in Delaunay elements. We then compute the fixed points in $e_1$ - $\omega_1$ space. In sec. \ref{s3} we compare our analytic results to direct three body integrations. We uncover the parameter space for librating, circulating and unstable orbit and its complex, fractal boundary. We also provide a numerical fit for the boundary between stable circulating and librating orbits. Finally in sec. \ref{s5} we conclude our results and discuss the limitations and future work.

\section{Modified fixed points} \label{s2}

\subsection{Hierarchical systems}\label{s2.1}

Consider a hierarchical system with masses $m_0,m_1 \ll m_2$, orbital separations $a_1 \ll a_2$ and other orbital elements denoted by subscript $1$ for the inner orbit and $2$ for the outer orbit. The double-averaged quadrupole Hamiltonian is \citep{naoz16}

\begin{align}
{\mathcal{H}}_{{\rm sec}} & \equiv\mathcal{\langle\langle H}\rangle\rangle=C\left[1-6e_{1}^{2}-3(1-e_{1}^{2})\cos^{2} \iota +15e_{1}^{2}\sin^{2} \iota \sin^{2}\omega_{1}\right]\label{eq:sec_quad}\nonumber \\
C & \equiv\frac{G\mu_{\rm in} m_2}{8a_{2} \left(1-e_{2}^{2}\right)^{3/2}}\left(\frac{a_{1}}{a_{2}}\right)^{2},
\end{align}
where $\mu_{\rm in}=m_0 m_1/(m_0+m_1)$ is the reduced mass of the inner binary, $\omega_1$ is the argument of pericentre and $\iota$ is the mutual inclination between the orbits.

We can change the variables into the Delauanay elements
\begin{align}
L_{1} & =\mu_{{\rm in}}\sqrt{G\left(m_{0}+m_{1}\right)a_{1}}\qquad L_{2}=\mu_{{\rm out}}\sqrt{G\left(m_{0}+m_{1}+m_{2}\right)a_{2}}\nonumber \\
G_{1} & =L_{1}\sqrt{1-e_{1}^{2}}\qquad G_{2}=L_{2}\sqrt{1-e_{2}^{2}}\nonumber \\
H_{1} & =G_{1}\cos i_{1}\qquad H_{2}=G_{2}\cos i_{2}\label{eq:delaunay}
\end{align}
where $\mu_{{\rm out}}=\left(m_{0}+m_{1}\right)m_{2}/(m_0+m_1+m_2)$
and $i_{i}$ are the inclinations with respect to the invariable plane.

Generally, the mutual inclination is given by $\cos \iota = \cos i_1 \cos i_2 + \sin i_i \sin i_2 \cos(\Omega_1 - \Omega_2) $ where $\Omega_i$ are the arguments of the ascending nodes. The Delaunay elements are strictly defined only on the invariable plane \citep[where the total angular momentum is parallel to the $\hat{\boldsymbol{z}}$ direction][]{sod75, naoz13}, where the nodes can be eliminated, $\Omega_2 - \Omega_1 = \pi$, and the mutual inclination reduces to $\cos \iota = \cos (i_1+i_2)$.

\subsubsection{Mild hierarchy}\label{s2.1.2}

The double averaging approach involves averaging over the inner, and then over the outer orbit. This approach ignores the relative movement of the outer body after one orbital period, and induces errors of the order of $P_1/P_2$. \cite{cb04} studied the corrections
 expanding the disturbing function in terms of $P_1/P_2$. \citep{luo16} generalised the analysis to general three body systems. The additional term in \cite{luo16} (their Eq. 39) is 
 \begin{align}
     {\mathcal{H}}_{\rm B}&=-\epsilon_{{\rm SA}}C\frac{27}{8}j_{z}\left(\frac{(1-j_{z}^{2})}{3}+8e_1^{2}-5e_{1}^{2}\sin^2\omega_1\sin^2\iota\right) + \mathcal{O}(e_2^2).
     \label{h_Brown} 
 \end{align}
Here,
\begin{equation}
    \epsilon_{\rm SA} = \left(\frac{a_1}{\ell_2}\right)^{3/2} \left( \frac{m_2^2}{(m_0+m_1)(m_0+m_1+m_2)} \right ) ^{1/2} = \frac{2P_2}{3\pi \tau_{\rm sec}} \label{eps_sa}
\end{equation}
 is the parameter that measures the relative accuracy of the double-averaging approximation,  and  $\ell_2 = a_2(1-e_2^2)$ is the semi-latus rectum. {The deviation between single-averaged and double-averaged approximations is proportional to $\epsilon_{\rm SA}$. In the limit of $e_2\ll 1$ and $m_2 \gg m_1, m_0$, we have $\epsilon_{\rm SA} = (m_2 a_1^3 / m_2 a_2^3)^{1/2}=P_1/P_2$, which is just $\epsilon$ from Eq. \ref{eq1} for the Lunar case.}
 
{In the right-hand side of Eq. \ref{eps_sa}, the secular timescale is given by }
\begin{equation}
\tau_{\rm sec} = \frac{L_1}{6C} = \frac{4(m_0+m_1)^{1/2}\ell_2^3}{3G^{1/2}m_2a_1^{3/2}}
\end{equation}
{Note that it differs by $4/3$ from the definition of \cite{luo16}. This is an order-of-magnitude estimate. Formally, one needs to integrate over the closed loop }
\begin{equation}
    \tau_{\rm sec} = \oint {\rm d}t = \oint \left(\frac{{\rm d}X} {{\rm d}t} \right)^{-1} {\rm d} X
\end{equation}
{Where $X \propto G_1^\alpha \propto (1-e_1^2)^{\alpha/2}$ is related to the action $G_1$. \cite{antognini15} used $\alpha=1$  in the test particle limit, while \cite{hamers-zlk} and \cite{ham19} used $\alpha=2$ for comparable masses for for a more general tidal galactic field, respectively.} 
 
 The Hamiltonian in Eq. \ref{h_Brown} has appeared in several different forms and is known as the Brown Hamiltonian \citep{brown1, brown2, brown3}. Recently, \citep{tremaine23} showed that the various forms in the literature are equivalent due to a gauge-freedom. Moreover, choosing the right gauge leads to a much simpler expression for the eccentric term. We keep the outer orbit circular to simplify the calculation.
 \begin{figure*}
    \centering
    \includegraphics[width=0.498\textwidth]{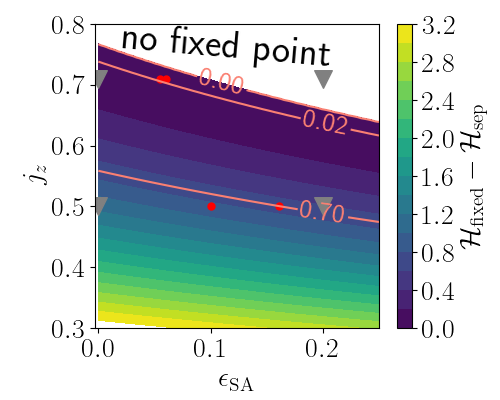}
    \includegraphics[width=0.498\textwidth]{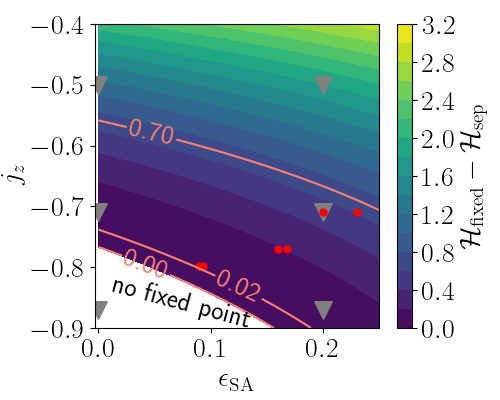}
    \caption{Values of the total Hamiltonian $\mathcal{H}_{\rm sec} + \mathcal{H}_B$ at the fixed point normalised at the separatrix. For each $\epsilon_{\rm SA}, j_z$ we plot $\mathcal{H}_{\rm fixed}=\mathcal{H}(e_{\rm fix}, \omega=\pi/2| \epsilon_{\rm SA}, j_z)$ minus the value at the separatrix (Eq. \ref{hs}). Left: Prograde orbits, $j_z>0$. Right: Retrograde orbits. The grey triangles are the initial conditions in Fig. \ref{fig1}, and \ref{fig2} where the red points are the initial conditions for Fig. \ref{fig4}, and \ref{fig5}.} \label{h_fixed}
\end{figure*}
The total Hamiltonian and the equations of motion can be written in terms of the Delaunay elements:

\begin{align}
\mathcal{H} & = \mathcal{H}_{\rm sec}(G_1, \omega_1, H_1, \Omega_1) + \mathcal{H}_{\rm B}(G_1, \omega_1, H_1, \Omega_1)\nonumber \\ & =  C \left[ 6\frac{G_1^2}{L_1^2} -3\frac{H_{1}^2}{L_{1}^{2}}+15\left(1-\frac{{G_{1}^{2}}}{L_{1}^{2}}-\frac{H_{1}^{2}}{G_{1}^{2}}+\frac{{H_{1}^{2}}}{L_{1}^{2}}\right)\sin^{2}\omega_1 \right] \nonumber \\ &  -\epsilon_{{\rm SA}}C\frac{27}{8}j_{z}\left[\frac{(1-j_{z}^{2})}{3}+8+(-8+5\sin^{2}\omega_1)\frac{G_{1}^{2}}{L_{1}^{2}}\right. \nonumber \\ & +\left.5\sin^{2}\omega_1\left(\frac{H_{1}^{2}}{G_{1}^{2}}-1-j_{z}^{2}\right) \right] + \mathcal{O}(e_2^2),   \label{htot}
\end{align}
{The first term gives as the standard ZLK equations of motion \citep[e.g.][]{lml15}. The second term gives us the additional contribution from Brown's Hamiltonian.}  For the argument of pericentre, we have
\begin{align}
\left. \frac{d\omega_1}{dt} \right|_{\rm B} & =\frac{d\mathcal{H}_{\rm B}}{dG_1}=-\epsilon_{{\rm SA}}C\frac{27}{8}j_{z}\left(-16\frac{G_{1}}{L_{1}^{2}}+10\sin^{2}\omega_1\left( \frac{G_{1}}{L_{1}^{2}}-\frac{H_1^2}{G_{1}^{3}} \right)  \right) \nonumber \\ 
 & =-\epsilon_{{\rm SA}}\frac{6C}{L_{1}\sqrt{1-e_{1}^{2}}}\frac{9}{8}j_{z}\left(-8(1-e_{1}^{2})+5\sin^{2}\omega_1(1-e_{1}^{2})\right. \nonumber \\ &-\left.5\cos^{2}\iota \sin^{2}\omega_1\right). \label{dwdt}
\end{align}
For a sanity check we will take $\omega_1=\pi/2$ and $e_1=0$ to get
\begin{equation}
    \tau_{\rm sec} \frac{d\omega_1}{dt} = 9 \epsilon_{{\rm SA}}\cos \iota \left(1-\frac{5}{8}\sin^{2} \iota \right).
\end{equation}

When the total (quadrupole and Brown) Hamiltonians are added, we get back Eq. 38 of \cite{cb04} (a prefactor of $3/4$ is missing in the first two terms, which appears in their Eq. 4). For almost coplanar orbit, $\sin \iota \approx 0$ and the typical apsidal advance rate is $\dot{\omega}_1=2+9\epsilon_{\rm SA} = 2(1 + 9\epsilon_{\rm SA}/2)$. Similarly, the nodal precession rate can also be derived from the Hamiltonian and is $\dot{\Omega}_1=-9\epsilon_{\rm SA}/32$, so the precession of the longitude of pericentre is $\dot{\varpi}_1 = 2 + (9+9/24)\epsilon_{\rm SA} = 2+ 225\epsilon_{\rm SA}/24$, which is compatible with earlier works on the Lunar theory (the secular timescale is defined up to a multiplicative factor of $3/4$). 

For completeness, the equation of motion for the eccentricity is 
\begin{align}
\left. \tau_{\rm sec} \frac{de_{1}}{dt}\right|_{\rm B} &=\tau_{\rm sec}\frac{\sqrt{1-e_{1}^{2}}}{L_{1}e_{1}}\frac{dG_{1}}{dt}=-\tau_{\rm sec}\frac{\sqrt{1-e_{1}^{2}}}{\tau_{\rm sec}L_{1}e_{1}}\frac{\partial\mathcal{H}_{{\rm B}}}{\partial\omega_{1}} \nonumber \\
 & = \frac{45}{64}\epsilon_{{\rm SA}}(1-e_{1}^{2})e_{1}\cos\iota\sin^{2}\iota\cos2\omega_{1}. \label{dedt}
\end{align}

Usually, the equations of motion are presented in terms of the vector elements (e.g. \citealp{luo16}). \cite{will21} derived the equations of motion in terms of the orbital elements directly from the Laplace-Largange secular equations (their Eq. 2.33). However, the derivative of $\omega_1$ is not clearly stated, but rather given by $\dot{\omega}_1 = \dot{\varpi}_1 - \dot{\Omega}_1\cos i_1$, where $\dot{\varpi}_1$ and $\dot{\Omega}_1$ are provided\footnote{Note that this is different from the usual definition of the longitude of pericentre $\varpi = \omega + \Omega$. Also eliminating $i_1$ is not trivial.}. \cite{sod75} also derived Brown's Hamiltonian in Delaunay variables, but presented the equations of motion only for small $e_1\ll 1$ (which is applicable for eclipsing binaries). 

Our equations reduce to the ones of \cite{sod75} for low $e_1$ and to our knowledge, this is the first time Eq. \ref{dwdt} was derived explicitly for any $e_1$.

\subsection{Fixed points}\label{s2.3}

{Restoring back to orbital elements in Eq. \ref{dwdt} and adding the ZLK secular contribution leads to}
\begin{align}
\dot{\omega}_1 & = 2(1-e_1^2) + 5\sin^2\omega_1(e_1^2 - \sin^2 \iota) \\ \nonumber
& - 9\epsilon_{\rm SA}j_z \left( -(1-e_1^2) + \frac{5}{8}\sin^2\omega_1(1-e_1^2) - \frac{5}{8}\cos^2 \iota \sin^2\omega_1 \right).
\end{align}
Setting $\omega_1=\pi/2$, and defining $x^2=1-e_1^2$, $\cos \iota = j_z / x^{1/2}$ we get 
\begin{align}
    0& =2x^2 - 5x^2+5\frac{j_z^2}{x^2}  - 9\epsilon_{\rm SA}j_z \left( -x^2 + \frac{5}{8}x^2 - \frac{5}{8}\frac{j_z^2}{x^2} \right)\nonumber \\
    &=  3\left(-1 + \frac{9}{8}\epsilon_{\rm SA} j_z \right)x^2 + 5\left(1 + \frac{9}{8}\epsilon_{\rm SA} j_z \right)\frac{j_z^2}{x^2},
\end{align}
and the solution is 
\begin{equation}
    x^2 = \sqrt{ \frac{5(1+\frac{9}{8}\epsilon_{\rm SA}j_z)}{3(1-\frac{9}{8}\epsilon_{\rm SA}j_z)} } |j_z|,
\end{equation}
or in terms of the eccentricity 
\begin{align}
    e_{\rm fix}=\sqrt{ 1 - \sqrt{ \frac{5(1+\frac{9}{8}\epsilon_{\rm SA}j_z)}{3(1-\frac{9}{8}\epsilon_{\rm SA}j_z)} } |j_z|}. \label{efix}
\end{align}

The associated inclination at the fixed point is
\begin{equation}
    \cos \iota_{\rm fix} = \frac{j_z}{x^{1/2}}= {\rm sign}(j_z) \left( \frac{3}{5} \right)^{1/4} \sqrt{\frac{1-\frac{9}{8}\epsilon_{\rm SA} j_z}{{1+\frac{9}{8}\epsilon_{\rm SA} j_z}} }|j_z|^{1/2}. \label{ifix}
\end{equation}

{Note that for the hierarchical limit ($\epsilon_{\rm SA}=0$), we get}
\begin{equation}
e_{\rm fix} = \sqrt{1 - \left(\frac{5}{3}\right)^{1/2}|j_z|},   
\end{equation}
and the associated inclination at the fixed point is
\begin{equation}
    \cos \iota_{\rm fix} = \frac{j_z}{x^{1/2}}={\rm sign}(j_z) \left(\frac{3}{5}\right)^{1/4}|j_z|^{1/2},
\end{equation}
which is the result of \cite{antognini15}. We note that analytic solutions have been obtained in \cite{kinoshita07} and the results for the general quadrupole problem (beyond the test particle limit) have been recently extended in \cite{hamers-zlk}.

\subsection{Phase space and individual orbits}

{The dynamics are revealed in more detail upon inspecting the total Hamiltonian $\mathcal{H}_{\rm sec} + \mathcal{H}_B$ from Eq. \ref{eq:sec_quad} and \ref{h_Brown}. The maximal value of the Hamiltonian is attained at the fixed point (if exists) for each value of $\epsilon_{\rm SA}$, $j_z$. The value itself is unimportant (Hamiltonians are always defined up to a constant), but an important quantity is the distance to the separatrix\footnote{Strictly speaking, a separatrix curve passes through an unstable (hyperbolic) fixed point, similar to the pendulum. The unstable fixed point is at the origin in rectangular coordinates $(e_1 \cos \omega_1, e_1 \sin \omega_1)$. The timescale also diverges logarithmically (e.g. Eq. 48 of \citealp{antognini15}).} -- a critical curve that separates librating and circulating orbits.}

 \begin{figure*}
    \centering
    \includegraphics[width=0.9\textwidth]{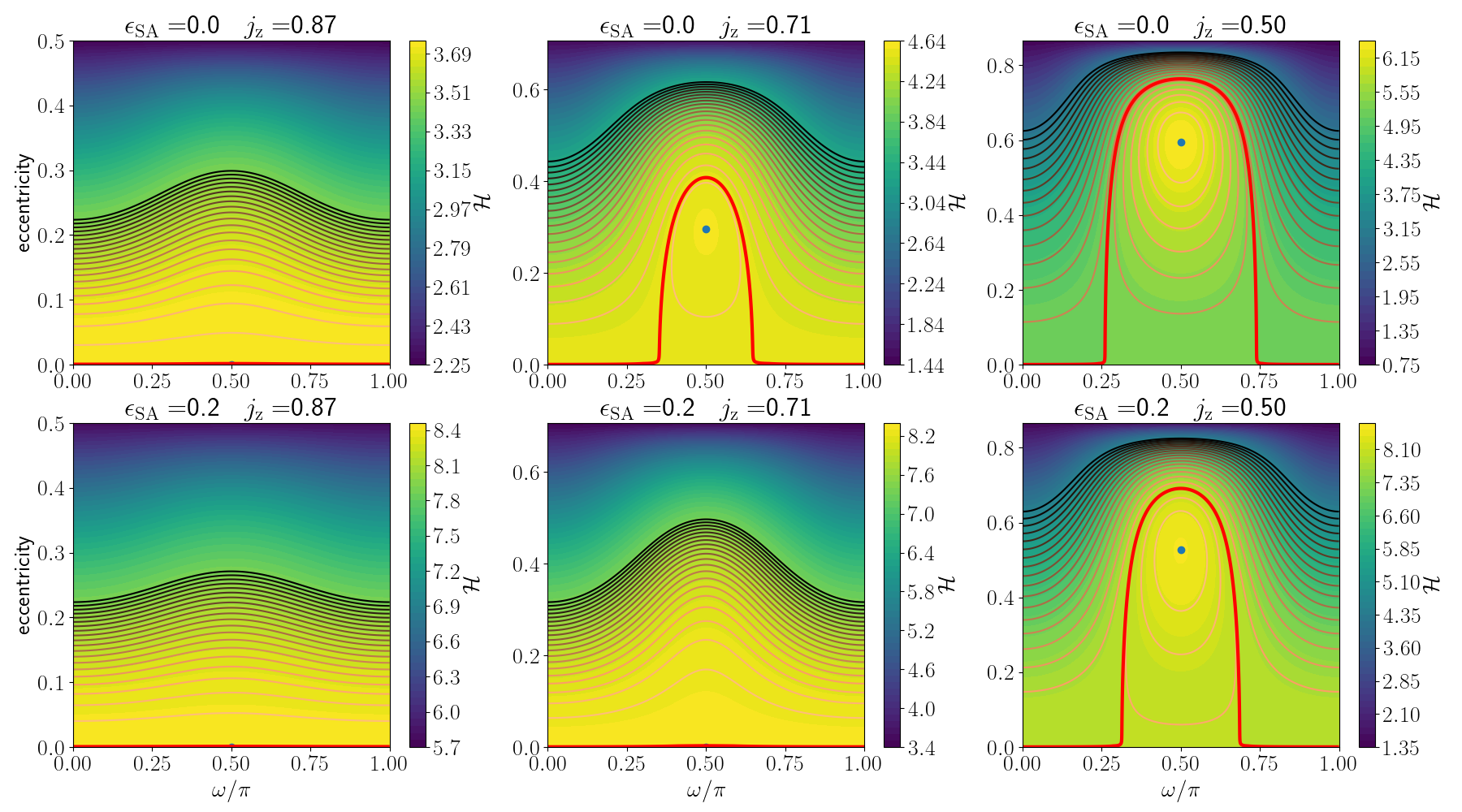}
    \caption{Phase space of the total Hamiltonian $\mathcal{H}_{\rm sec} + \mathcal{H}_{B}$ for the prograde cases. Top panels show only $\mathcal{H}_{\rm sec}$, while the lower panel have both terms with $\epsilon_{\rm SA}=0.2$. The colour and lines show constant energy curves. The thick red line is the separatrix which separates circulating and librating orbits. {The Hamiltonian is normalised by $C=1$}.}\label{fig1}
\end{figure*}
\begin{figure*}
    \centering
    \includegraphics[width=0.9\textwidth]{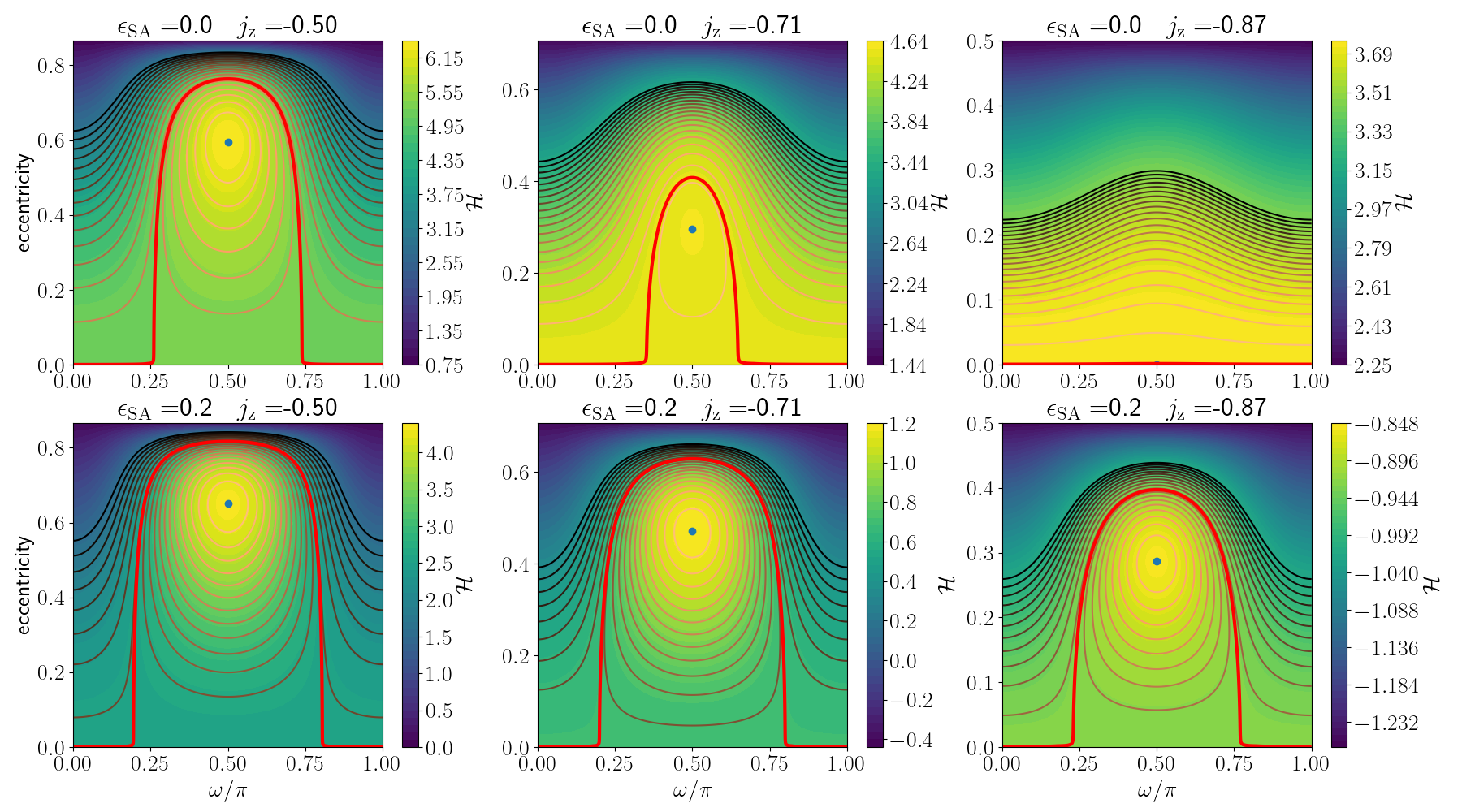}
    \caption{Same as Fig. \ref{fig1}, but for retrograde orbits.} \label{fig2}
\end{figure*}
{Fig. \ref{h_fixed} shows the distance of the maximal value $\mathcal{H}_{\rm fixed}=\mathcal{H}(e_{\rm fix}, \omega=\pi/2| \epsilon_{\rm SA}, j_z)$ from the separatrix}
\begin{equation}
    \mathcal{H}_{\rm sep} = 1 - 3j_z^2 - \frac{9\epsilon_{\rm SA}}{8}j_z(1-j_z^2), \label{hs}
\end{equation}
{as function of $\epsilon_{\rm SA}$, and $j_z$. We see that the trend in the prograde orbits (left) is different from the trend in the retrograde orbits (right), where the distance is decreasing (increasing) with larger $\epsilon_{\rm SA}$ for prograde (retrograde) orbits. This is another manifestation of the symmetry breaking that occurs when introducing the Brown Hamiltonian \citep[e.g.][]{i79,grishin2017,gpf18}. For decreasing $|j_z|$, further away from the bifurcation point, the distance is increasing, as expected. Very low $|j_z|$ are expected to have the largest distance}

{Although it is tempting to try and determine whether an orbit will librate or circulate, the situation is more complex, especially for larger $\epsilon_{\rm SA}$. The modified averaging procedure that is used to obtain $\mathcal{H}_B$ leaves out higher order terms and osculating elements. In other words, the quantities $j_z$, $\epsilon_{\rm SA}$ are conserved only on secular timescales, but fluctuate on shorter tiemscales. One can decompose, e.g. $j_z$ to mean and fluctuating contributions, $\bar{j}_z + \delta j_z$. The mean $\bar{j}_z$ is conserved because $\mathcal{H}_{\rm sec} + \mathcal{H}_B$ is independent of the nodal angle $\Omega_1$. The fluctuating elements vary with period $P_2$. Although an analytic envelope for the strength of the amplitude of $\delta j_z$ can be estimated (e.g. \citealp{luo16}), it is challenging to draw definite conclusions on the nature of the orbits without exact N-body integrations. Nevertheless, we suggest a procedure that finds an empirical criterion for librating orbits that relies on the difference $\mathcal{H}_{\rm fixed} - \mathcal{H}_{\rm sep}$ and the analytic envelope of $\delta j_z$ and apply it to irregular satellites in paper II \citep{gri24II}.}

Fig. \ref{fig1} shows the constant energy curves in the phase space for prograde orbits ($j_z>0$) with and without the Brown Hamiltonian. {The Hamiltonian is normalised by $C=1$.} The numerically obtained fixed point is consistent with our Eq. \ref{efix} (blue dots). The thick red line is the separatrix.

Fig. \ref{fig2} shows the constant energy curves for retrograde orbits. We see that increasing $\epsilon_{\rm SA}$ could lead to 'birth' of fixed points, contrary to the prograde case where an existing fixed point may be eliminated. 

{In choosing the initial conditions in Fig \ref{fig1} and \ref{fig2} we show how the morphology of the phase space changes once the correction from the Brown Hamiltonian are introduced. We see, for example, that the general volume of the librating zone is decreasing for prograde orbit (or the fixed point completely vanishes) for increasing $\epsilon_{\rm SA}$. On the contrary, retrograde orbits ar increasing their librating volume phase space with increasing $\epsilon_{\rm SA}$ (or a fixed point and libration zone can be born). The minimal value of the Hamiltonian can be obtained analytically but is of little importance. Large negative value will result in almost co-planar motion on almost constant eccentricity. For a given $j_z$ the maximum eccentricity allowed is $\sqrt{1-j_z^2}$ on a circular orbit. }

\begin{figure*}
    \centering
    \includegraphics[width=0.95\textwidth]{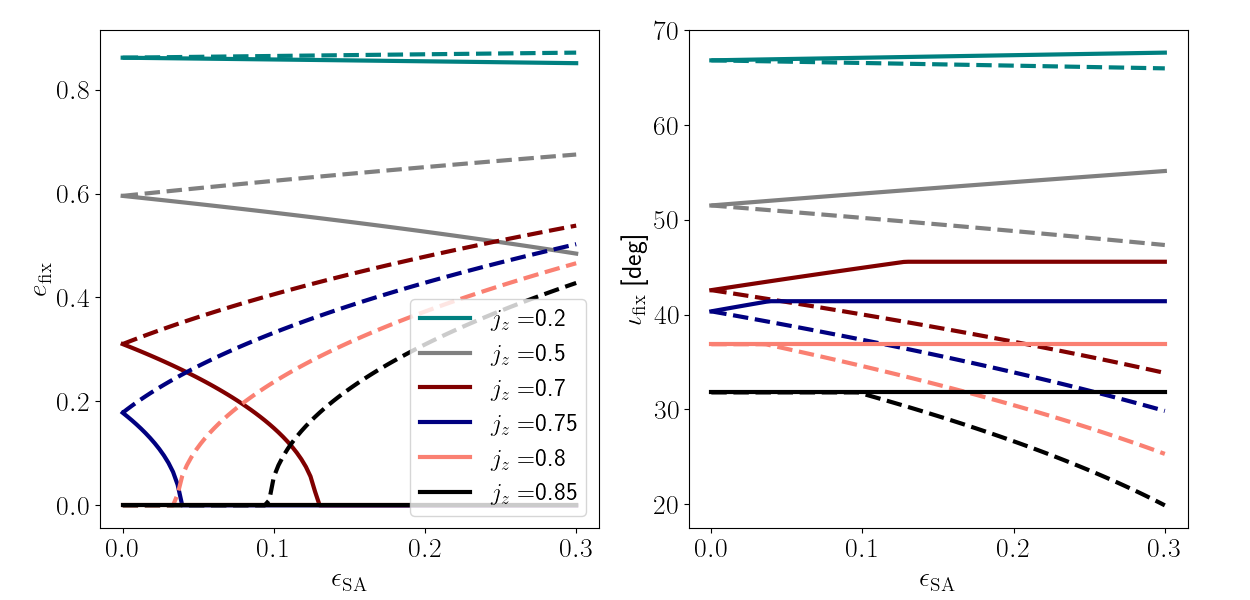}
    \caption{Left: Eccentricity of the fixed point as a function of $\epsilon_{\rm SA}$ for various $j_z$ values. Right: The associated inclination at the fixed point. The dashed lines correspond to retrograde orbits where $j_z \to -j_z$. The retrograde inclination is understood as $\iota_{\rm fix}^{\rm retro} = 180^{\circ} - \iota_{\rm fix}$. } \label{fig3}
\end{figure*}

Fig. \ref{fig3} shows the eccentricity (left) and associated inclination (right) of the fixed point as a function of $\epsilon_{\rm SA}$, for various values of $j_z$. The retrograde orbits are dashed where $j_z \to - j_z$ and $\iota_{\rm fix} - 180^{\circ} - \iota_{\rm fix}$. We see that the eccentricity decreases for prograde orbits and increases for retrograde ones. This asymmetry stems from the Coriolis asymmetry \citep{i80} and comes from odd powers of $j_z \propto \cos \iota$, which also affects the maximal eccentricity and critical inclination for ZLK oscillations \citep{gpf18}. For low $j_z$ (both positive and negative) the changes are mild, which is due to the scaling of $1-e_{\rm fix}$ with $j_z$, so it doesn't affect too much the highly eccentric orbits (but even the correction from Brown Hamiltonian could be broken down if $\sqrt{1-e_{\rm max}^2} \lesssim \epsilon_{\rm SA}^2$, see \citealp{gpf18} for details).

To summarise, the addition of the Brown Hamiltonian {induce bifurcations which change the morphology of the phase space for fixed $j_z$, and varying $\epsilon_{\rm SA}$ leading to birth and death of ZLK resonance:} For prograde orbits, with increasing $\epsilon_{\rm SA}$, the fixed point could vanish (e.g. for $j_z=0.7$, maroon line), while for retrograde orbits the fixed point could be in areas which are inaccessible with quadrupole approximation only (e.g. dashed black line).

\section{Comparison to N-body integrations}\label{s3}

{We've seen that the deviation from the double-averaging approximation linearly increases with $\epsilon_{\rm SA}$. Moreover}, systems with larger $\epsilon_{\rm SA}$ also tend to be more unstable: If an inner orbit is parametrised in terms of the Hill radius, $a_1=kr_{\rm H} = k a_2 ((m_0+m_1)/3m_2)^{1/3}$, then $\epsilon_{\rm SA} = 3^{-1/2}k^{3/2}$. The critical Hill stability limit is around $k\approx 0.5$ for prograde orbits \citep{grishin2017,tory22}, so the maximal $\epsilon_{\rm SA} \approx 0.2$. 

Even if the system is dynamically stable, mildly hierarchical systems also exhibit additional fluctuations, both in osculating elements on shorter timescales and higher-order corrections to the Brown Hamiltonian. {Therefore, the averaged}, constant energy $\mathcal{H}_{\rm sec} + \mathcal{H}_B$ { is subjected to short-term fluctuations due to the osculating elements \citep[see e.g. sec. 3.1 of][]{luo16}}. The amplitude of the osculating elements is also proportional to $\epsilon_{\rm SA}$. At some point, $\epsilon_{\rm SA}$ is large enough such that an orbit that starts near the fixed point could cross the separatrix and become circulating. 
\begin{figure*}
    \centering
    \includegraphics[width=0.49\textwidth]{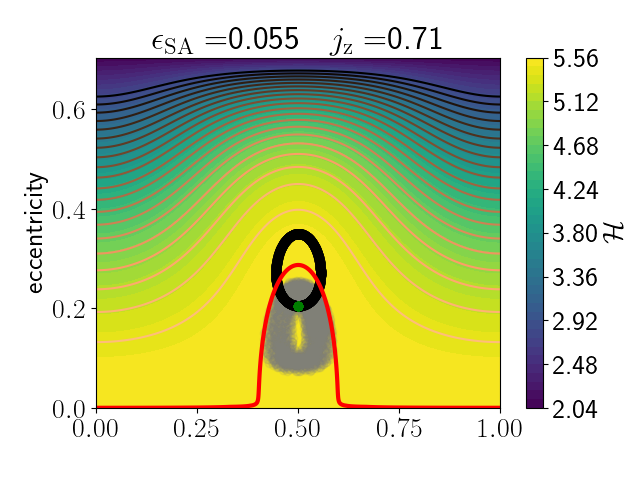}
    \includegraphics[width=0.49\textwidth]{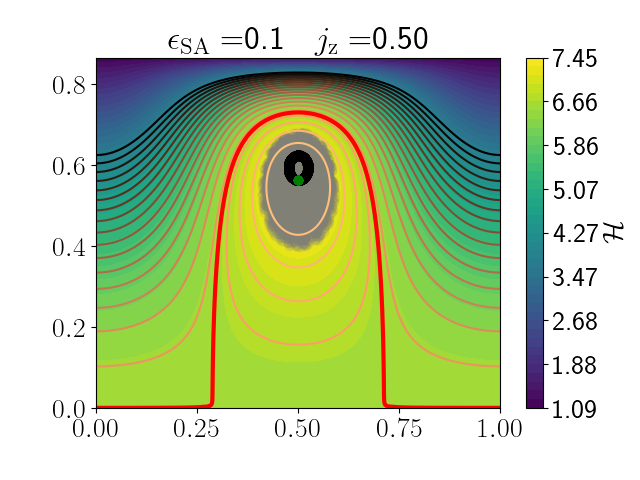}   
    \includegraphics[width=0.49\textwidth]{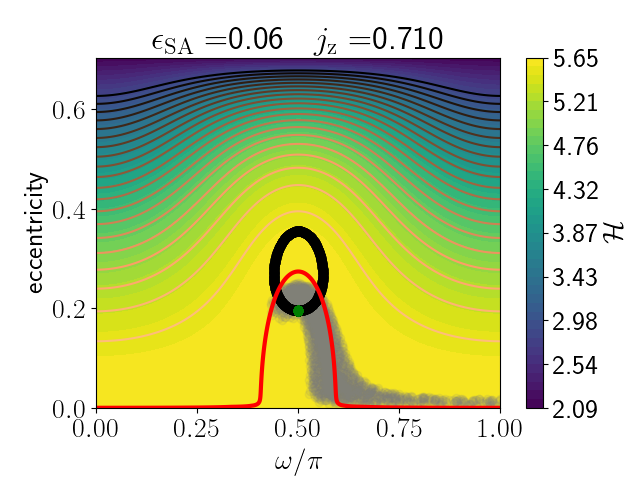}
    \includegraphics[width=0.49\textwidth]{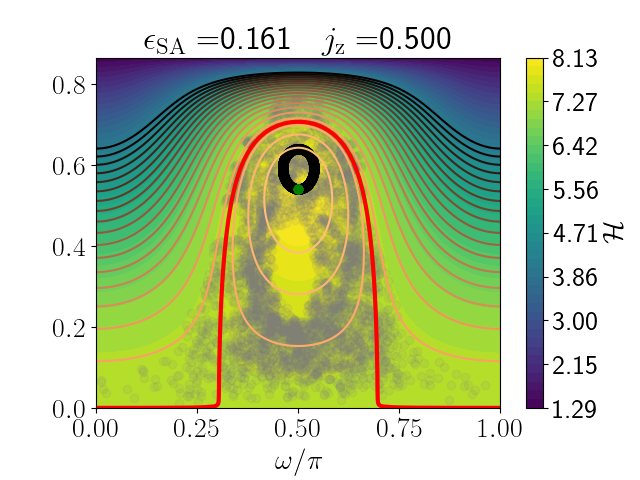}   
 
    \label{prograde-nbody}

    \caption{Comparison to N-body simulations for the prograde case. The grey points are snapshots of {one} orbit at different times, which starts at the fixed point. {Black points are results of secular evolution with $\mathcal{H}_{\rm sec}$, the green points is the secular evolution with $\mathcal{H}_{\rm B}$, which remains fixed.} The top panels show librating orbits, while the bottom panels show orbits that escape and circulating. The colours and lines are the same as in Fig. \ref{fig1} and \ref{fig2}.} \label{fig4}
\end{figure*}
\begin{figure*}
    \centering
     \includegraphics[width=0.33\textwidth]{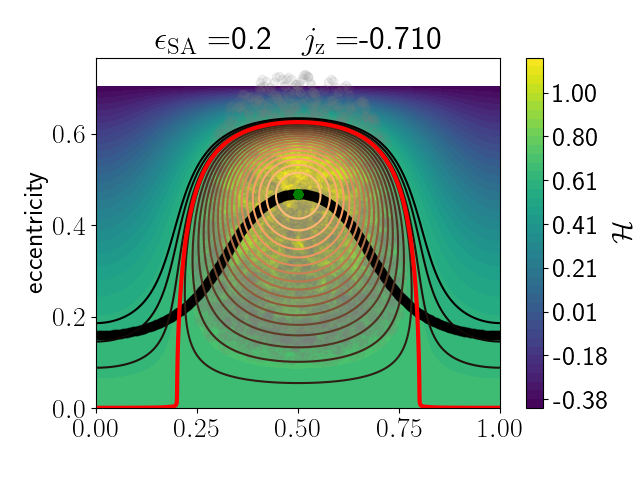}
     \includegraphics[width=0.33\textwidth]{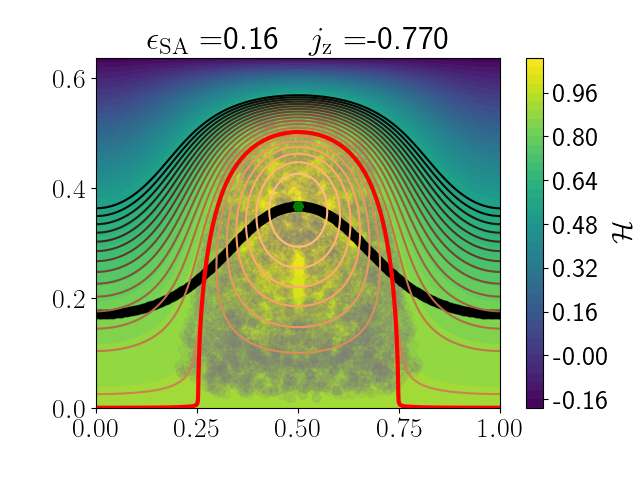}
     \includegraphics[width=0.33\textwidth]{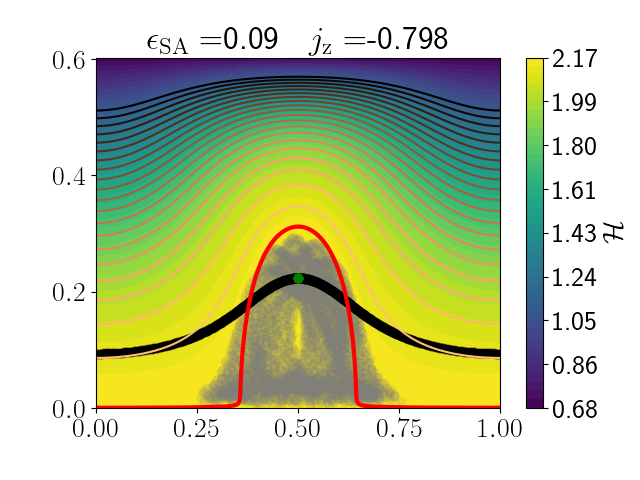}
    \includegraphics[width=0.33\textwidth]{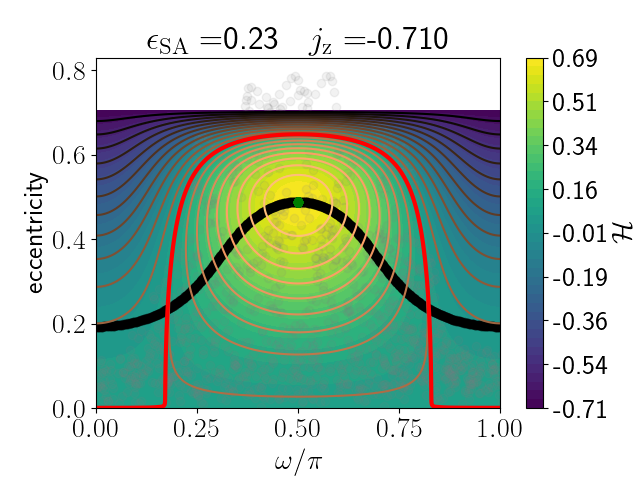}
     \includegraphics[width=0.33\textwidth]{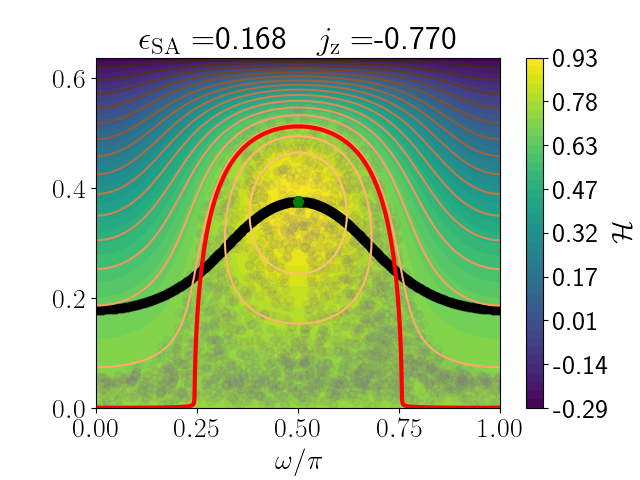}
     \includegraphics[width=0.33\textwidth]{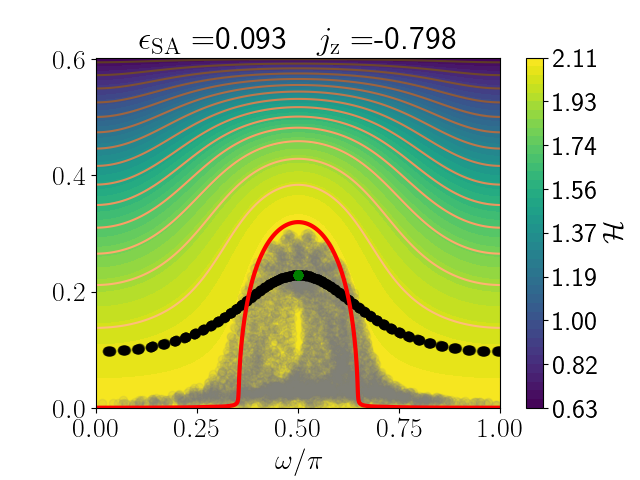}

    \caption{Similar to Fig. \ref{fig4} but with retrograde orbits.}     \label{fig5}
\end{figure*}

\begin{figure*}
    \centering
   \includegraphics[width=0.99\textwidth]{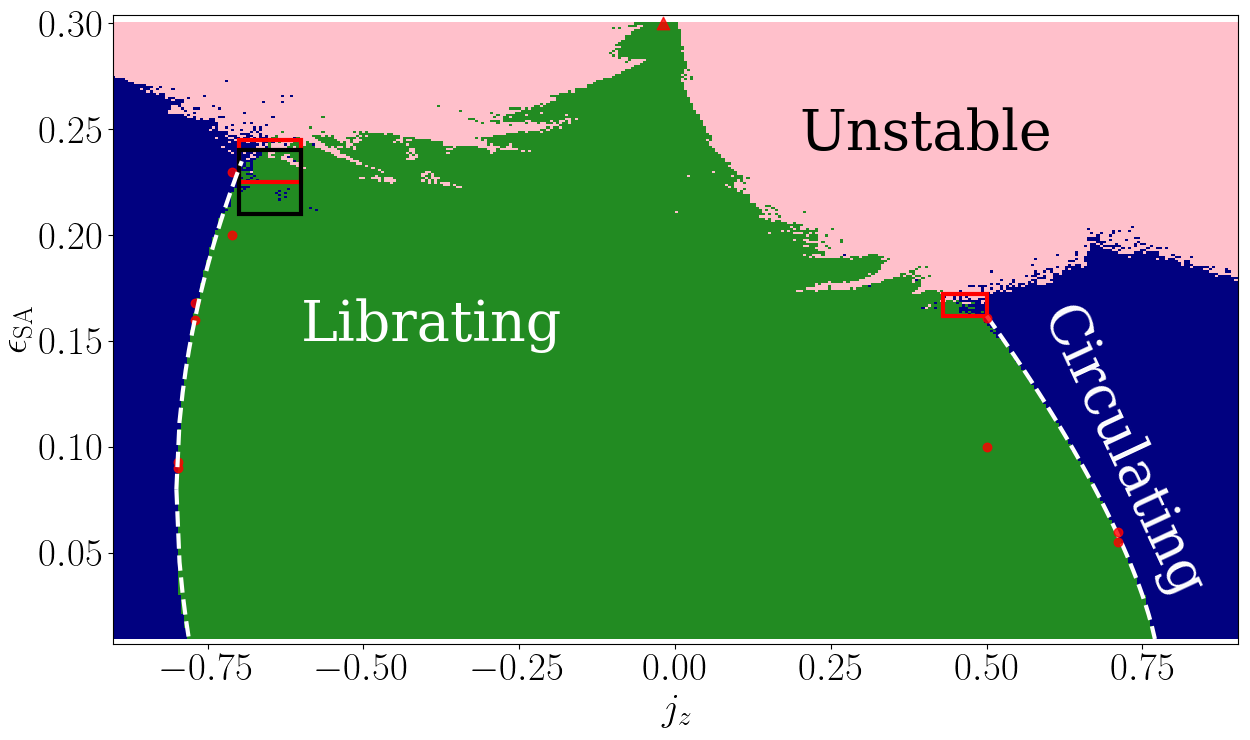}
   
    \caption{Phase map of N-body integration. The plot include $360$ initial values of $j_z  \in [-0.9, 0.9]$ and $291$ values of $\epsilon_{\rm SA} \in [0.01,0.3]$, a total of $104,760$ integrations. The end time is $t_{\rm end} =n_{\rm c}P_{\rm out} \epsilon_{\rm SA}^{-1}$, {where $n_{\rm c}=10$ for $\epsilon_{\rm SA}<0.15$ and $n_{\rm c}=100$ if $\epsilon_{\rm SA} \ge 0.15$}. This ensures that dozens of secular timescales are integrated, but also saves integration time for the more stable systems. Green dots are librating orbits, navy are circulating orbits, and pink are unstable orbits. The dashed white lines are the numerical fits of Eq. \ref{eps_fit}. {Red points are individual integration shown in Fig. \ref{fig4} and \ref{fig5}. Red triangle is the orbit in Fig. \ref{B1}. Boxed are areas of high-resolution further insets: see Fig. \ref{fig7} for the red insets and Fig. \ref{fig6.3} for the black inset with longer integration time of $n_{\rm c}=1000$.} }
    \label{fig6}
\end{figure*}
\begin{figure*}
    \centering

   \includegraphics[width=0.47\textwidth]{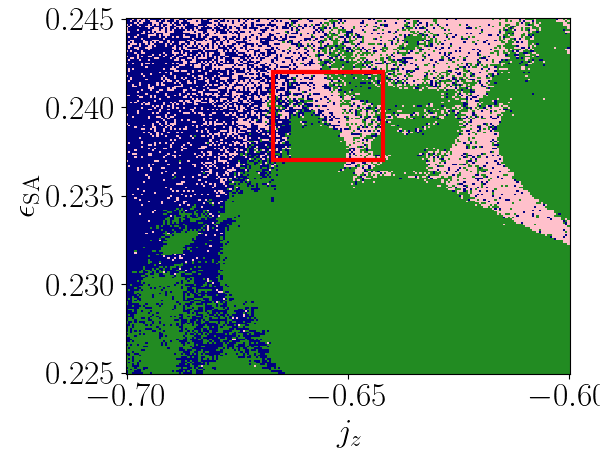}
   \includegraphics[width=0.47\textwidth]{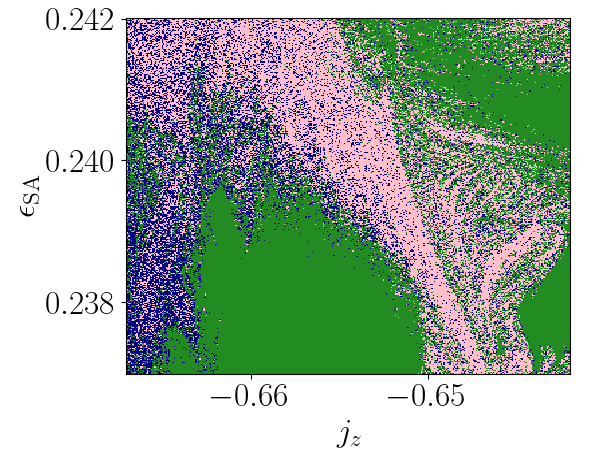}
   \includegraphics[width=0.47\textwidth]{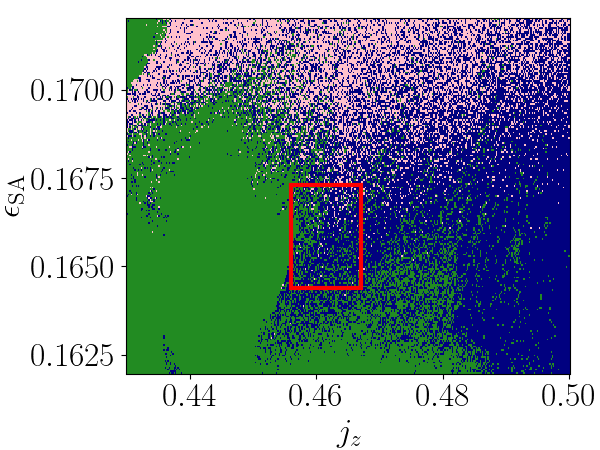}
   \includegraphics[width=0.47\textwidth]{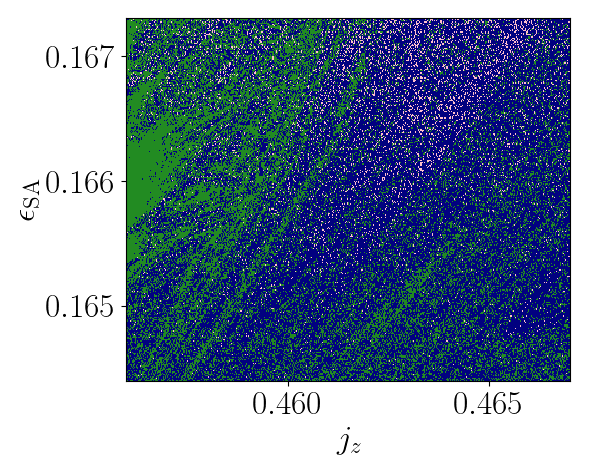}
   
    \caption{ High resolution insets on the boundaries. Top row: firsts insets, with $201\times201=40,401$ ($201\times 351=70,551$) integrations for the retrograde (prograde) inset in the top (bottom) left. Bottom row: additional inset of the left panels with $331\times331=109,591$ ($291\times441=128,331$) initial conditions.}
    \label{fig7}
\end{figure*}
\begin{figure}
    \centering
     \includegraphics[width=0.47\textwidth]{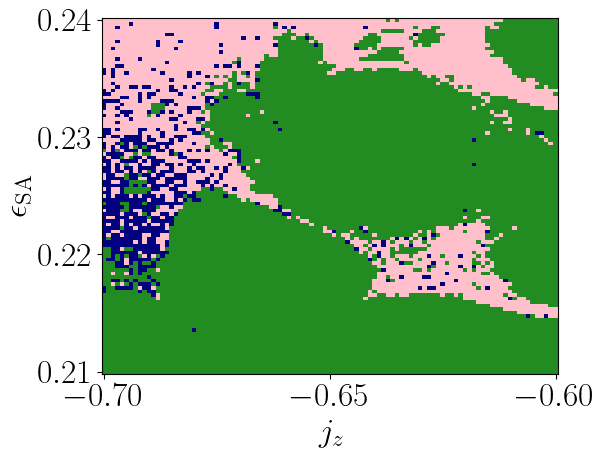}
    \caption{Inset of Fig. \ref{fig6} at higher resolution of $101\times 101=10201$ orbits integrated for longer times of $1000n_{\rm c}$.}     \label{fig6.3}
\end{figure}
\subsection{Individual orbits}

In order to explore the critical transition to circulating orbits, we compare the results of sec. \ref{s2} with direct N-body integrations. All the integrations were computed using the \href{https://rebound.readthedocs.io}{REBOUND} N-body code, with the IAS15 adaptive timestep integrator, accurate to machine precision for billions of orbits \citep{ias15}. 

Unless stated otherwise, for all orbits we vary $\epsilon_{\rm SA}$ and $j_z$ to get the fixed points' solution for $e_{\rm fix},\iota_{\rm fix},\omega_1=\pi/2$ and $a_1$ which corresponds to $\epsilon_{\rm SA}$. The other orbital elements are for the outer binary are $a_2=5.203\ \rm au$, $m_2=m_J$ (all other angles are zero). We keep the outer orbit circular for simplicity. For the inner orbit, the mean anomaly is $\pi$ to start the orbit at apocentre, and $\Omega_1=\pi/4$, to ensure that the osculating $j_z$ is equal to its averaged value \citep{gpf18}. {The interaction time is $t_{\rm end} = 10P_{\rm out} \epsilon_{\rm SA}^{-1}$, which ensures that the simulations are integrated to about $\mathcal{O}(10)$ secular timescales.}

\subsubsection{Prograde orbits}
In Fig. \ref{fig4} we show similar energy contours as in Fig. \ref{fig1}, but with the grey dots as recorded from direct three body integrations for various initial conditions. {To clarify, each panel has only one orbit. The osculating orbital elements tend to fill up a fraction of the phase space, but the averaged value of the energy is retained.} The recorded orbital elements are osculating elements, which fluctuate about the size of $\epsilon_{\rm SA}$ around the mean orbital elements \citep{luo16}. 

{We've also ran the same initial conditions with the secular code \texttt{SecuLab} \citep{gp22}, which includes a variant of the Brown Hamiltonian and solves the equations of motion from \cite{luo16} but for arbitrary reference frame. The black points are where $\mathcal{H}_{\rm B}$ is turned off ($\epsilon_{\rm SA}=0$), which does not fit within the separatrix in the left panels. The green points are the results of the secular evolution, which remains fixed.}

We see that for $j_z=0.71$ the critical $\epsilon_{\rm SA}$ for which the orbit escapes is $\epsilon_{\rm SA} \approx 0.07$, while for $j_z=0.5$ it is around $\epsilon_{\rm SA} \approx 0.16$. The associated inclinations for the top panel from Eq. \ref{ifix} are $45.02\ \rm deg$ and $53.97\ \rm deg$ for $j_z=0.71$ and $j_z=0.5$, respectively, and for the bottom panel  $45.47\ \rm deg$ and $55.345\ \rm deg$ for $j_z=0.71$ and $j_z=0.5$, respectively.

\subsubsection{Retrograde orbits}

Fig. \ref{fig5} shows the numerical integrations for retrograde orbits, where $j_z=-0.71, -0.77, -0.798$. We see that the retrograde orbits are more stable, with larger values of $\epsilon_{\rm SA}$ allowed. This is consistent with the stability analysis \citep{grishin2017, tory22}, however, the latter studies have generally assumed an initially circulating orbit with low eccentricity.  The associated inclinations for the top panel from Eq. \ref{ifix} are $147.49\ \rm deg, 152.61\ \rm deg$ and $147.79\ \rm deg$ for $j_z=-0.71, -0.77$ and $j_z=-0.8$, respectively, and for the bottom panel  $153.26\ \rm deg, 153.4\ \rm deg$ and $148.62\ \rm deg$  for  $j_z=-0.71, -0.77$ and $j_z=-0.8$, respectively. {We also see that the secular approximation completely missed out the libration in all the cases.}

{In Fig. \ref{fig4} and \ref{fig5} we show how a small increase in $\epsilon_{\rm SA}$ leads to a qualitative change in the N-body integration. Our analytic theory is insufficient to describe the validity of the existence of fixed point and it requires detailed numerical investigation, which is done in the next section. We note that individual time evolution of various quantities for several irregular satellites, and the boundaries for the osculating energy, are shown in paper II of this series \citep{gri24II}.}

\subsection{Parameter space exploration}

In order to qualitatively explore the possibility of librating orbits at large $\epsilon_{\rm SA}$ values, we integrate a large sample of orbits. We uniformly initialise $361$ values of $j_z \in [-0.9, 0.9]$ and $291$ values of $\epsilon_{\rm SA} \in [0.01,3]$ and set an orbit at the fixed point in $e_{\rm fix}, \iota_{\rm fix}$ given by Eq. \ref{efix} and \ref{ifix} and $\omega_1=\pi/2$. {If the fixed point doesn't exist, the initial eccentricity will be zero (and the orbit can only be circulating or unstable). The timescale for each orbit is $t_{\rm end}=n_{\rm c}P_{\rm out} \epsilon_{\rm SA}^{-1}$, where $n_{\rm c} = 10$ for $\epsilon_{\rm SA} <0.15$, and $n_{\rm c} = 100$, for $\epsilon_{\rm SA} \ge 0.15.$. This ensures that at least several secular timescales are integrated and longer timescales are tested. For lower $\epsilon_{\rm SA}$, values, $n_{\rm c}$ is lower to ensure reasonable integration times. The entire grid takes $\sim 2$ days to run on a 8-core machine.} The planet is initialised in a circular orbit at a separation of $5.203\ \rm au$ from the Sun to mimic Jupiter. We deem the orbit unstable if  $d>3a_1$ where $d$ is the instantaneous separation. We classify the orbits as circulating if at any point $\sin\omega_1<-0.2$ (since librating orbits will be constrained to $\omega_1 \in (0,\pi)$, or $\sin \omega_1>0$). {Choosing $-0.2$ is a compromise between making sure that librating orbits are not misclassified and reducing the run time of a system that completes a full circulation.}

Fig. \ref{fig6} shows the result of the integration. We see that most of the orbits are able to remain in the librating zone. The transition in the retrograde case is between $-0.8 \le j_z \le -0.75$, while in the prograde case the transition is more dependent on $\epsilon_{\rm SA}$, varying from $j_z \sim 0.5$ for large $\epsilon_{\rm SA}$ to $j_z \sim 0.75$ for small $\epsilon_{\rm SA}$. For stable regions, we can find numerically the transition between the librating and circulating orbits. We find the following numerical fit
\begin{equation}
    \epsilon_{{\rm SA}}(j_{z})=\begin{cases}
0.4(\sqrt{3/5}-j_{z})^{0.7} & j_{z}>0;\ \epsilon_{{\rm SA}}\in[0.01,0.16]\\
0.08+0.5(0.8+j_{z})^{0.52} & j_{z}<0;\ \epsilon_{{\rm SA}}\in[0.08,0.23]\\
0.08-0.54(0.8+j_{z})^{0.52} & j_{z}<0;\ \epsilon_{{\rm SA}}\in[0.01,0.08]
\end{cases}, \label{eps_fit}
\end{equation}
which is plotted in dashed white lines in Fig. \ref{fig6}. Note that the retrograde fit {could be} multi-valued {in $\epsilon_{\rm SA}$ for fixed $j_z$,} and the minimal $j_z$ s attained at $\epsilon_{\rm SA} \approx 0.08$. 

{We also plot additional insets of Fig. \ref{fig6} with higher resolution. Fig. \ref{fig7} shows higher resolution grids of the red rectangles. An additional second zoom-in is presented in the right panels. We see that higher resolution revealed additional structure, which indicates a fractal boundary.}

{Fig. \ref{fig6.3} shows the insets of the black rectangle, run for longer times of $1000n_{\rm c}$. We see that the longer integration times make more orbits unstable, but most of the librating orbits (green) are retained. A Large number of the circulating orbit is becoming unstable. We also checked the stability of a few initially circulating orbits (e.g. $\epsilon_{\rm SA}=0.22, j_z=-0.69$), which become unstable after $n_c \approx 20,000$ cycles. We suspect that the transition to circulation could ba a pathway to long-term instability for orbits near the stability limit. Further simulations with longer integration times are required to explore this transition and will be studied elsewhere.}
\begin{figure*}
    \centering

   \includegraphics[width=0.99\textwidth]{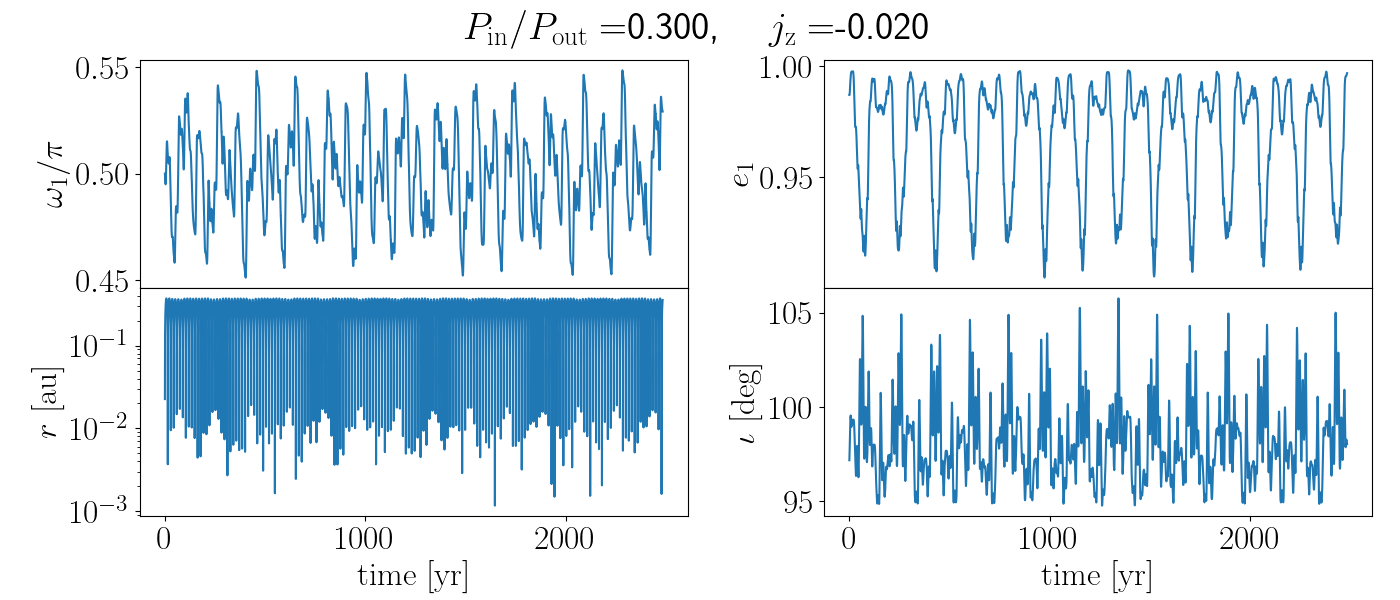}

    \caption{Highly inclined orbital integration. The period ratio, or $\epsilon_{\rm SA} = 0.3$. The different panels show the evolution of the argument of pericentre $\omega_1$, eccentricity $e_1$, instantaneous separation $r$ and mutual inclination $\iota$. }
    \label{B1}
\end{figure*}

{Another interesting feature is the stable, highly inclined orbits around $\epsilon_{\rm SA} \approx 0.3$.  We Show in Fig. \ref{B1} the orbital evolution of a particular orbit with $\epsilon_{\rm SA}=0.3, j_z=-0.02$. we integrated the orbit for $n_c=100$ secular cycles. We also ran a longer integration for $n_c=10^5$ secular cycles and verified that the orbit remained stable.}

{Although the orbit is chaotic, the argument of pericentre is librating and stable. The reason may lie in the fact that $\mathcal{H}_{\rm fixed} - \mathcal{H}_{\rm sep}$ is larger for lower vales of $j_z$ (see the trend in Fig. \ref{fig1}). Thus, even highly perturbed orbits could retain a libration motion of $\omega_1$. Additional work is required to tackle these orbits. In practice, the large radial extent (left bottom panel) will cause a satellite to either collide with the planet or to be scattered with other satellites, so it is likely to be lost either via collision or ejection.}

\section{Summary and conclusions}\label{s5}
We studied analytically and numerically the long-term evolution of triple systems of mild hierarchy, where the period ratio between the inner and outer orbits is not too small. Contrary to other studies, we focused on the fixed points of the Brown Hamiltonian in eccentricity $e_1$ and argument of pericentre $\omega_1$ phase space. Our conclusions are summarised below:

\begin{itemize}
 
\item We have calculated for the first time the fixed points of the Brown Hamiltonian (Eq. \ref{efix}-\ref{ifix} and Fig. \ref{fig3}). The fixed point are expressed in terms of dimensionless quantities $j_z=\sqrt{1-e_1^2}\cos \iota$ and $\epsilon_{\rm SA}$ (Eq. \ref{eps_sa}), and could be applied to wide range of astrophysical applications.

\item We've integrated numerically several orbits to test the limitation of our analytic approach and visualise the morphology of the orbits (Fig. \ref{fig4}-\ref{fig5}. We found that the retrograde librating orbits extend to large values of $|j_z|$ and $\epsilon_{\rm SA}$, and for marginal orbits the escape usually occurs in the low eccentricity stage.

\item We have numerically explored the boundary between circulating, librating and unstable orbits (Fig. \ref{fig6}-\ref{fig7}). We found a complex, fractal boundary. Retrograde orbits tend to be more stable (a known result for initially circular orbits, \citealp{grishin2017, tory22}) and librating retrograde orbits tend to occupy a larger {region of} parameter space. We provide a numerical fit for the librating-circulating boundary $\epsilon_{\rm SA}(j_z)$ in Eq. \ref{eps_fit}, and find that the critical $j_z$ has a larger range for prograde orbits, while retrograde librate can be recorded both for larger $|j_z|$ and $\epsilon_{\rm SA}$. 
\end{itemize}

Future work will expand the Brown Hamiltonian formalism beyond the test particle limit where the three masses are comparable, and/or where the outer orbit is eccentric. Also, the exact transition to circulating orbit and the timescale for transition could also behave chaotically and are interesting future research directions. 

Extra apsidal advance from other, non-Keplerian forces may quench the ZLK cycles. Considering additional non-Keplerian forces (either conservative or dissipative) will be relevant for other applications in stellar and compact object dynamics. For example, general-relativity and tidal evolution could affect the long-term evolution and the final fate of triple systems. 

In paper II we focus on the long-term evolution of irregular satellites in the Solar system. We will construct a condition on the energy of the satellite from Brown's Hamiltonian and the analytic fluctuating terms. We'll confirm that the known librating satellites satisfy our criterion, and find several new librating satellites.

\section*{Acknowledgements}
EG thanks Alessandro Trani, Isobel Romero-Shaw, Mor Rozner, Rosemary Mardling and Scott Tremaine for stimulating discussions and comments on the manuscript. EG acknowledges support from  ARC grant FT190100574 (CI: Mandel).

\section*{Data Availability}
No data used in this work. The methods are publicly available in the \href{https://github.com/eugeneg88/fixed_points_brown}{\texttt{GitHub} repository}.

\section*{Appendix A: Libration-circulations boundary}
\begin{figure*}
    \centering

   \includegraphics[width=0.49\textwidth]{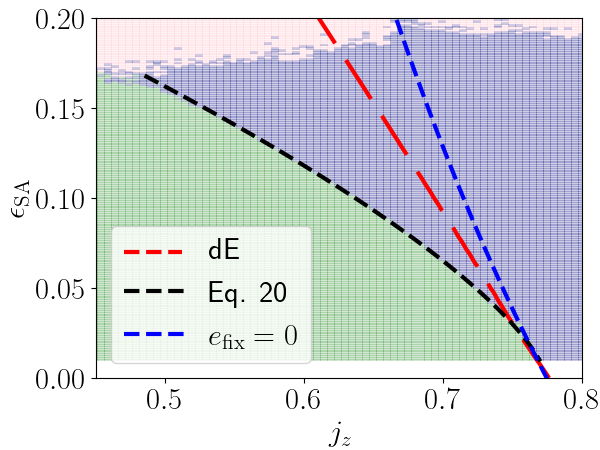}
   \includegraphics[width=0.49\textwidth]{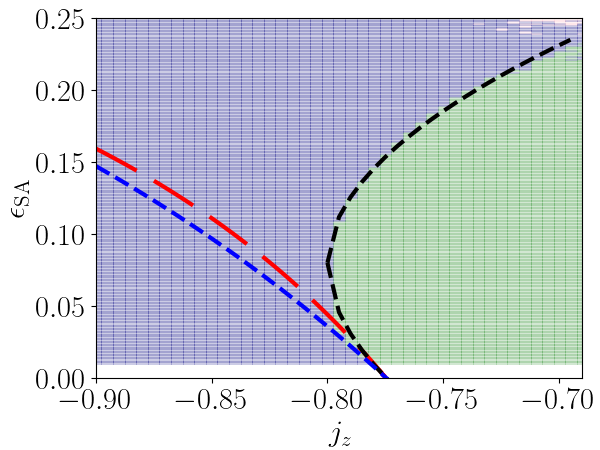}

    \caption{ Additional semi-analytical to fit the librating-circulating binary, zoomed in on the prograde boundary (left) and the retrograde boundary (right). In addition to the dashed black line from Eq. \ref{eps_fit} (originally dashed white in Fig. \ref{fig6}), we plot also a fit based on setting $e_{\rm fix}=0$ (dashed blue), and on the energy fluctuations (dashed red). See text fro details. The transparent green and blue grid have the same meaning as in Fig. \ref{fig6} and \ref{fig7}. }
    \label{A1}
\end{figure*}

In Fig. \ref{fig6} we used an empirical fit based on numerical integration (Eq. \ref{eps_fit}). Here we attempt two additional semi-analytic fits: 

i) Bifurcation point: We take $e_{\rm fix}=0$ from Eq. \ref{efix}. This yields the relation
\begin{equation}
    \epsilon_{\rm SA} = \frac{8}{9j_z}\frac{1-\frac{5}{3}j_z^2}{1+\frac{5}{3}j_z^2}
\end{equation}

ii) Energy fluctuation: We vary the total Hamiltonian at the fixed point $\mathcal{H}_{\rm fixed}$ to $d\mathcal{H}_{\rm fixed}=\mathcal{H}(e_{\rm fix}, \omega=\pi/2| \epsilon_{\rm SA}, j_z \pm \Delta j_z)$. Here, $\Delta j_z$ is the typical $j_z$ energy fluctuation which is proportional to $\epsilon_{\rm SA}$ (see details and exact expressions \citealp{luo16} and paper II). We pick the sign in $j_z \pm \Delta j_z$ which gives the smaller $d\mathcal{H}_{\rm fixed}$. We then plot the the contour where  $d\mathcal{H}_{\rm fixed}$ equals the energy at separatrix.

Fig. \ref{A1} shows the various fits with the numerical data. We see that all the fits converge for $\epsilon_{\rm SA} = 0$ to $j_z=\sqrt{3/5}\approx0.775$ and also have the same first order expansion of $\epsilon_{\rm SA} (j_z)$. However, the curved are markedly diverge for $\epsilon_{\rm SA} \gtrsim 0.5$. The energy condition (ii) performs slightly better, but quickly diverges too (dashed red lines), while the bifurcation point (i) is ever further (dashed blue lines) from the empirical fit. Although the general trend is correct, the analytical fits tend to diverge for larger $\epsilon_{\rm SA}$, which requires further investigation. 

A condition that uses a variant of the energy condition is developed in paper II and applied to the available data on irregular satellites.

\bibliography{sample631}
\bibliographystyle{mnras}

\bsp	
\label{lastpage}
\end{document}